\documentclass[amsmath,superscriptaddress,prb,twocolumn]{revtex4-2}
\usepackage{bm}
\usepackage{bbm}
\usepackage{enumerate}
\usepackage{graphicx}
\usepackage[dvips]{epsfig}
\usepackage{epsf}
\usepackage{physics}
\usepackage{xcolor}
\usepackage{bbold}
\usepackage[normalem]{ulem}
\usepackage[caption=false]{subfig}
\usepackage[colorlinks=true,allcolors=blue]{hyperref}

\makeatletter
\newcommand{\Rmnum}[1]{\expandafter\@slowromancap\romannumeral #1@}
\makeatletter

\allowdisplaybreaks

\begin{document}

\title{Spin-wave dynamics controlled by tunable ac magnonic crystal}

\author{Ankang Liu}
\affiliation{Department of Physics and Astronomy, Texas A\&M University, College Station, Texas 77843-4242, USA}
\author{Alexander M. Finkel'stein}
\affiliation{Department of Physics and Astronomy, Texas A\&M University, College Station, Texas 77843-4242, USA}
\affiliation{Department of Condensed Matter Physics, The Weizmann Institute of Science, Rehovot 76100, Israel}

\begin{abstract} 
The magnonic crystal, which has a spatial modulation wave vector $q$, couples the spin wave with wave vector $k$ to the one with wave vector $k-q$. For a conventional magnonic crystal with direct current (dc) supply, the spin waves around $q/2$ are resonantly coupled to the waves near $-q/2$, and a band gap is opened at $k=\pm q/2$. If instead of the dc current the magnonic crystal is supplied with an alternating current (ac), then the band gap is \emph{shifted} to $k$ satisfying $|\omega_{s}(k)-\omega_{s}(k-q)|=\omega_{ac}$; here $\omega_{s}(k)$ is the dispersion of the spin wave, while $\omega_{ac}$ is the frequency of the ac modulation. The resulting gap in the case of the ac magnonic crystal is the half of the one caused by the dc with the same amplitude of modulation. The time evolution of the resonantly coupled spin waves controlled by properly suited ac pulses can be well interpreted as the motion on a Bloch sphere. The tunability of the ac magnonic crystal broadens the perspective of spin-wave computing.
\end{abstract}

\maketitle

\noindent\emph{Introduction.} Interference of classical waves can be exploited for building computing devices that may work faster than the classical ones for certain tasks \cite{lloyd1999quantum,knight2000quantum,meyer2000sophisticated,ferry2001quantum,patel2006optimal}. As it was demonstrated in Ref. \cite{balynsky2021quantum}, a computing machine relying on the spin-wave interference outperforms a classical digital computer on a database search. In that work, the authors utilized a linear superposition of two spin waves with different phases to encode a ``qubit'' state, with each ``qubit'' sent through its own waveguide and controlled by an individual phase shifter. However, scaling up such a spin-wave computing device will be a significant challenge. To address this challenge, in this paper we explore the possibilities of using a single \emph{tunable} ac magnonic crystal for the control of different spin-wave pairs.

Spin waves are the collective wave excitations in the magnetically ordered system, which have the frequencies typically ranged from GHz up to even THz. The utilization of spin waves for the purposes of quantum-information exchange was proposed long ago \cite{khitun2001spin}. The later studies \cite{serga2010yig,collet2017spinwave} showed that the spin waves excited in yttrium iron garnet (YIG) can have both lifetime and coherence time longer than $100$ ns. In addition, the ability of exciting exchange spin waves with wavelengths down to the nanometer range \cite{che2020efficient,wang2022deeply} allows one to build compact spin-wave-based devices at the submicron scale. All these properties make the low-damping coherent spin waves a suitable candidate for performing rapid data processing and wave computing \cite{chumak2015magnon,csaba2017perspectives,mahmoud2020introduction,pirro2021advances,yuan2022quantum,chumak2022roadmap}.

\begin{figure}[htp] \centerline{\includegraphics[clip, width=1 \columnwidth]{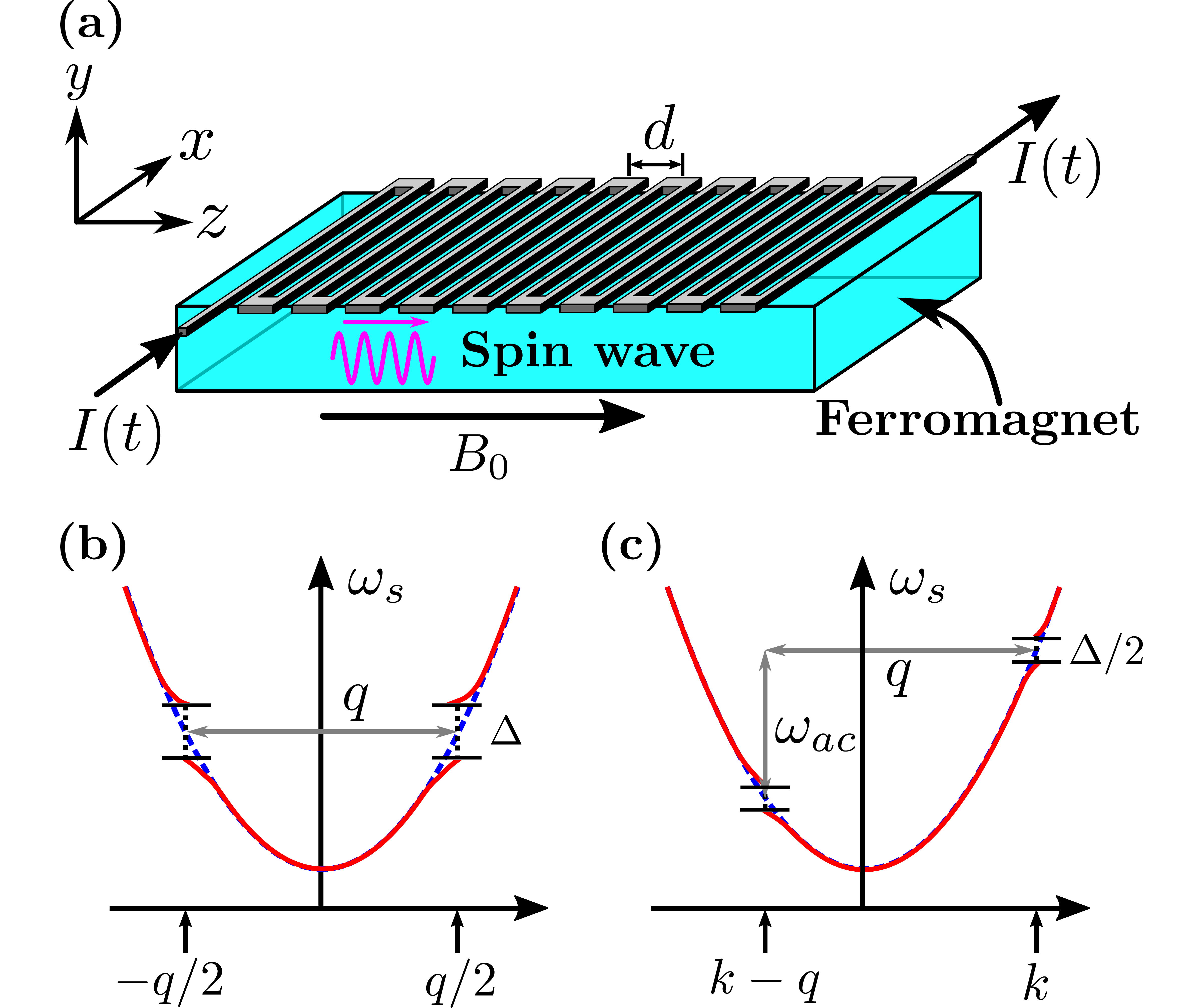}}
	
	\protect\caption{(a) A schematic setup of the current-induced magnonic crystal, which was used for studying the spin-wave dynamics in experiments \cite{chumak2010all,karenowska2012oscillatory}. The meander structure at the top of the ferromagnet creates a spatially modulated magnetic field, which is $\propto I(t)\cos(qz)\bm{e}_z$, along the $z$ direction. Here, $q=2\pi/d$. In the second row, we sketch the spin-wave spectrum when the magnonic crystal is switched on (the solid red curves) and when it is not effective (the dashed blue curves). Panel (b) is for the dc magnonic crystal while (c) is for the ac case (the gaps around the wave vectors $q-k$ and $-k$ are not shown). The band gap $\Delta\propto I_{0}$. Note that the band gap caused by the ac magnonic crystal is the half of the one created by the dc with the same amplitude $I_{0}$.}
	
	\label{fig:magnonic_crystal}
\end{figure}

As an efficient way to control spin waves, magnonic crystals have been studied both experimentally and theoretically \cite{chumak2010reverse,chumak2010all,karenowska2012oscillatory,chumak2017magnonic,liu2022control}. A prototypical current-induced magnonic crystal for ferromagnet is schematically depicted in Fig. \ref{fig:magnonic_crystal}(a) (cf. Refs. \cite{chumak2010all,karenowska2012oscillatory}). If a dc current, $I(t)=I_{0}$, is supplied through a metallic meander structure with a period $d$, a spatially modulated static magnetic field generates a magnonic crystal for the spin waves propagating along the direction of the modulation. For the spin waves with a symmetric spectrum, the dc magnonic crystal resonantly couples the spin waves with wave vectors around $\pm q/2$, where $q=2\pi/d$ is determined by the period of the spatial modulation. As shown in Fig. \ref{fig:magnonic_crystal}(b), the coupling of the two degenerate waves opens a band gap $\Delta$, which is $\propto I_{0}$ \cite{chumak2010all,karenowska2012oscillatory}. In practice, for an incident spin-wave packet, after switching on the dc magnonic crystal, the spin-wave components that are under the resonant scattering conditions start to alternate between the forward- and backward-propagating states; while the out-of-resonance spectral components are unaffected by the perturbation and propagate unidirectionally.

In this paper, we consider the same experimental setup as was exploited in Refs. \cite{chumak2010all,karenowska2012oscillatory}, but extend the discussion to an ac modulated magnonic crystal, i.e., when the current $I(t)=I_{0}\cos(\omega_{ac}t+\varphi_{ac})$. We show that under the limit $\omega_{ac}\gg\Delta$ a spin-wave pair with wave vectors $k$ and $k-q$ can be controlled by a \emph{tunable} ac magnonic crystal that satisfies the \emph{shifted resonance} condition $|\omega_{s}(k)-\omega_{s}(k-q)|=\omega_{ac}$ [cf. Fig. \ref{fig:magnonic_crystal}(c)].

\noindent\emph{Spin-wave scattering induced by magnonic crystal.} Suppose that, initially (i.e., at $t<0$), there was a free spin wave with wave vector $\bm{k}=k\bm{e}_z$ propagating inside the magnet. We consider a device fabricated from a ferromagnet with all spins located on a cubic lattice. To present the idea we will restrict ourselves to the spin-wave excitations originating from the short-range exchange couplings. This is sufficient for illustrating the concept of the shifted resonance. Note that the scheme proposed in this paper is general and applicable to all types of spin waves (e.g., the dipolar spin-wave modes). In the continuum limit, the spin operators become a space- and time-dependent variable $\bm{S}(\bm{r},t)$ (see Supplemental Material (SM) \cite{SM}\nocite{maekawa2017spin,holthaus2015floquet,oka2019floquet,cohen2020quantum,ndsolve,visscher1991fast} for the details). Before the magnonic crystal is switched on, the constant external magnetic field $\bm{B}=B_{0}\bm{e}_z$ aligns all spins along the $z$ direction in the ground state. In the case of a spin-wave excitation, $\bm{S}(\bm{r},t)$ deviates from the equilibrium and acquires small $S^{x,y}(\bm{r},t)$. The linearized equation in $S^{x,y}(\bm{r},t)$ can be solved by a plane wave $S^{+}(\bm{r},t)\equiv S^x(\bm{r},t)+iS^y(\bm{r},t)=(\Delta S)e^{i[\bm{k}\cdot\bm{r}-\omega_{s}(\bm{k})t+\varphi_s]}$ \cite{SM}, where $\Delta S$ is the amplitude of the spin wave, $\varphi_s$ is its initial phase, and $\omega_{s}(\bm{k})=Ak^2+\gamma B_{0}$ gives the spin-wave dispersion. Here, $A>0$ is determined by the ferromagnetic exchange coupling between the nearest neighboring spins and $\gamma$ is the gyromagnetic ratio of the spin.

Next, at $t=0$, one switches on the ac modulated magnonic crystal, which for $t>0$ is described by $\Delta B_{0}\cos(\omega_{ac}t+\varphi_{ac})\cos(qz)\bm{e}_z$. Here, $\Delta B_{0}$ is the intensity of the magnonic crystal controlled by $I_{0}$, the frequency of the ac modulation is given by $\omega_{ac}$, while $\varphi_{ac}$ is the initial phase determined at the moment $t=0$. In the discussed geometry, the spin-wave propagation is effectively one-dimensional. To find what will be the dynamics of the spin wave after $t=0$, one needs to solve $S^{+}(z,t)$ from the equation \cite{SM}
\begin{align}\label{EOM_Mp_time_h}
	\frac{dS^{+}}{dt}=&iA\nabla^2S^{+}-i\gamma[B_{0}
	\nonumber
	\\
	&+\Delta B_{0}\cos(\omega_{ac}t+\varphi_{ac})\cos(qz)]S^{+},
\end{align}
by matching the solution for $t>0$ with the free spin-wave solution for $t<0$.

The magnonic crystal term in Eq. (\ref{EOM_Mp_time_h}) couples the spin-wave state $k$ to the state $k-q$ as long as $|\omega_{s}(k)-\omega_{s}(k-q)|\approx\omega_{ac}$. To better understand the spin-wave dynamics under the ac magnonic crystal, we look for the solution of Eq. (\ref{EOM_Mp_time_h}) in the form
\begin{align}\label{ansatz_S_plus}
	S^{+}(z,t)=&(\Delta S)\big[\mathcal{S}_{p}(t)\sin\left(k_{+}z\right)+\mathcal{S}_{m}(t)\sin\left(k_{-}z\right)
	\nonumber
	\\
	&+\mathcal{C}_{p}(t)\cos\left(k_{+}z\right)+\mathcal{C}_{m}(t)\cos\left(k_{-}z\right)\big].
\end{align}
Here, we introduced four complex time-dependent coefficients $\mathcal{S}_{p/m}(t)$ and $\mathcal{C}_{p/m}(t)$ in front of the basis functions $\sin(k_{\pm}z)$ and $\cos(k_{\pm}z)$, respectively. Solution (\ref{ansatz_S_plus}) describes the mutual scatterings between a pair of the spin waves with the wave vectors $k=k_{+}=q/2+\delta k$ and $k-q=-k_{-}=-q/2+\delta k$. Although the scattering occurs between the spin waves with the oppositely directed wave vectors, the wave vectors $k_{\pm}$ are defined here as their absolute values and, therefore, are positive (we assume that $-q/2<\delta k<q/2$). Note that, in Eq. (\ref{ansatz_S_plus}), the contributions from the spin waves with wave vectors $\pm3q/2+\delta k$ and higher are neglected.

\begin{figure}[htp] \centerline{\includegraphics[clip, width=1 \columnwidth]{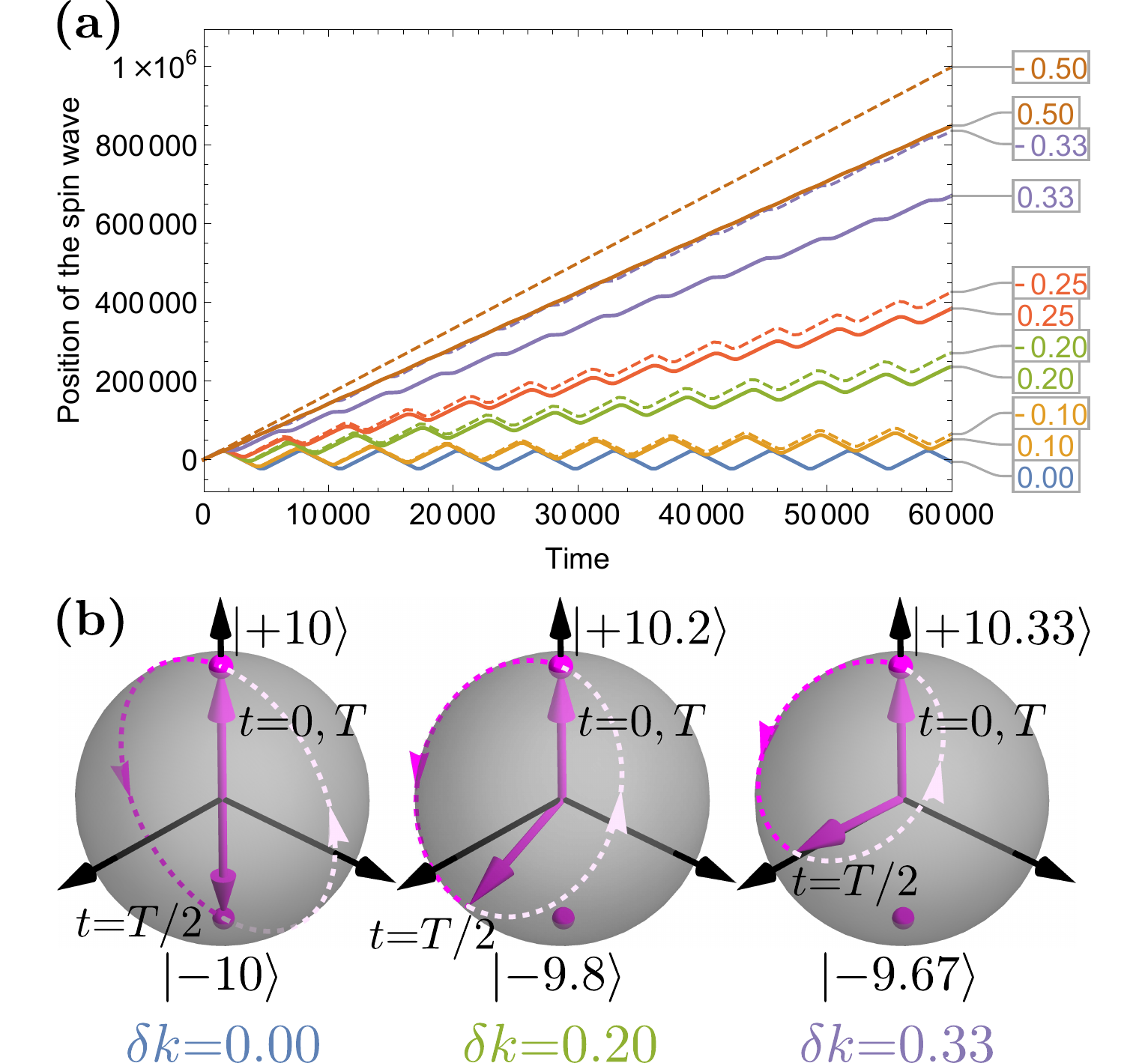}}
	
	\protect\caption{(a) Time-dependent position of the spin waves with wave vectors $q/2+\delta k$ after the dc magnonic crystal is switched on at $t=0$. Different $\delta k$ [in units of $(2\pi/1000)$] are indicated by the number in the rectangular boxes. Other parameters are $A=2$, $q=20\times(2\pi/1000)$, $\gamma B_{0}=1$, $\Delta=\gamma\Delta B_{0}=0.001$, and $\varphi_s=\varphi_{ac}=0$. The resonant solution that is closest to the horizontal axis is at $\delta k=0$. In our plots, the length on the vertical axis is in units of the lattice constant, which is taken to be $1$, while the time on the horizontal axis is measured in units $1/\gamma B_{0}$. The connection between the scales of the spin-wave dynamics in a real physical system and those shown in the figures is discussed in (iii) below. (b) The Bloch-sphere trajectories of the spin waves with three different $\delta k$. Here, the trajectories (dashed magenta curves with arrowheads on the surface of the sphere) are plotted only for the first period of the to-and-fro motion; the paths of the second half of the period are indicated by the light color. The intermediate positions of each of the waves are shown by the colored 3D arrows at $t=0$, $T/2$, and $T$, where the period $T=2\pi/\omega_{\downarrow\uparrow}(\xi)$.}
	
	\label{fig:to_and_fro_vs_BS_DC}
\end{figure}
\noindent\emph{Spin-wave dynamics in the dc case.} In the dc case, by using the ansatz (\ref{ansatz_S_plus}) and with the proper initial conditions, one gets a simple analytical solution \cite{SM}:
\begin{align}\label{sol_Mx_static}
	&S^{x}(z,t)=\Re[S^{+}(z,t)]=
	\nonumber
	\\
	&\Delta S\bigg[(1)\cos(\frac{\omega_{\downarrow\uparrow}t}{2})\cos\left(k_{+}z-\Omega t+\varphi_s\right)
	\nonumber
	\\
	&+\beta_1(\xi)\sin(\frac{\omega_{\downarrow\uparrow}t}{2})\sin\left(k_{+}z-\Omega t+\varphi_s\right)
	\nonumber
	\\
	&-\beta_2(\xi)\sin(\frac{\omega_{\downarrow\uparrow}t}{2})\sin\left(k_{-}z+\Omega t-\varphi_s\right)\bigg]
\end{align}
with two $\beta$ functions:
\begin{align}\label{betas}
	\beta_1(\xi)=\frac{\xi}{\sqrt{1+\xi^2}},\ \beta_2(\xi)=\frac{1}{\sqrt{1+\xi^2}}.
\end{align}
[The solution for $S^{y}(z,t)$, which has a similar structure, is presented in SM.] In Eq. (\ref{sol_Mx_static}), $\omega_{\downarrow\uparrow}\equiv\omega_{\downarrow\uparrow}(\xi)=(\gamma\Delta B_{0})\sqrt{1+\xi^{2}}$. Here $\xi$ is a dimensionless variable defined as $\xi(\delta k)\equiv\Delta\omega_{s}(\delta{k})/\gamma\Delta B_{0}$ with $\Delta\omega_{s}(\delta k)\equiv\omega_{s}(k_{+})-\omega_{s}(k_{-})$; note that $\Delta\omega_{s}$ depends on $\delta k$ critically. The parameter $\xi$ describes the level of the energy mismatch of the two spin waves participating in the scattering induced by the dc magnonic crystal; $\Omega\equiv\Omega(\delta k)\equiv[\omega_{s}(k_{+})+\omega_{s}(k_{-})]/2$ is a sort of ``central frequency" which depends on $\delta k$ non-critically. The spin-wave dynamics described by Eq. (\ref{sol_Mx_static}) reminds the one of the Rabi oscillations \cite{SM}.

In Fig. \ref{fig:to_and_fro_vs_BS_DC}(a) we plot a momentary ``position'' of the spin wave with wave vector $q/2+\delta k$ as a function of time for different $\delta k$ after the dc magnonic crystal is switched on at $t=0$. [The ``position'' was determined by tracking numerically the most left zero-crossing point of the imaginary part of the spin-wave solution (\ref{ansatz_S_plus}) within a sufficiently large spatial interval.] As one can see from Fig. \ref{fig:to_and_fro_vs_BS_DC}(a), at $\delta k=0$, the dependence of spin-wave position on time is a zigzag curve around the horizontal axis which indicates the resulting to-and-fro propagation of the spin wave. The period of this zigzag curve is extracted to be $T_{\downarrow\uparrow}\approx6283\approx2\pi/\gamma\Delta B_{0}$, which indicates that the gap $\Delta$ induced by the dc magnonic crystal is equal to $\gamma\Delta B_{0}$. Moreover, from Fig. \ref{fig:to_and_fro_vs_BS_DC}(a), one may conclude that the to-and-fro motion exists only for a limited interval of $\delta k$ when $|\delta k|<\delta k_c\approx0.33$ in units of $(2\pi/1000)$. The critical value $\delta k_c$ is roughly determined by the criterion $\Delta\omega_{s}(\delta k_c)=\gamma\Delta B_{0}$. For the chosen parameters, we get $\Delta\omega_{s}(0.33)\approx0.00104$ while $\gamma\Delta B_{0}=0.001$.

\noindent\emph{Spin-wave dynamics in the ac case.} Now, we turn to the ac modulated magnonic crystal. When $\omega_{ac}\neq0$, the general form of the solution pairs $\mathcal{S}_{p/m}(t)$ and $\mathcal{C}_{p/m}(t)$ is governed by Floquet theorem \cite{grifoni1998driven,bukov2015universal}, and the time dependence of the spin-wave dynamics is determined by the quasi-energies of the system. Here we focus only on an important limit when $\omega_{ac}\approx\Delta\omega_{s}\gg\gamma\Delta B_{0}$. In this limit, one can implement the rotating wave approximation (RWA) \cite{grifoni1998driven}. Under the RWA, we find the approximate solution \cite{SM}
\begin{align}\label{sol_Mx_time_depen}
	&S^{x}(z,t)=\Re[S^{+}(z,t)]\approx
	\nonumber
	\\
	&\Delta S\bigg\{(1)\cos(\frac{\omega_{R}t}{2})\cos\left[k_{+}z-\left(\Omega+\frac{\omega_{ac}}{2}\right)t+\varphi_s\right]
	\nonumber
	\\
	&+\beta_1(\tilde{\xi})\sin(\frac{\omega_{R}t}{2})\sin\left[k_{+}z-\left(\Omega+\frac{\omega_{ac}}{2}\right)t+\varphi_s\right]
	\nonumber
	\\
	&-\beta_2(\tilde{\xi})\sin(\frac{\omega_{R}t}{2})\sin\left[k_{-}z+\left(\Omega-\frac{\omega_{ac}}{2}\right)t-\varphi_s-\varphi_{ac}\right]\bigg\}
\end{align}
with $\omega_{R}\equiv\omega_{R}(\tilde{\xi})=(\gamma\Delta B_{0}/2)\sqrt{1+\tilde{\xi}^2}$, while $S^{y}(z,t)$ can be found from the imaginary part of $S^{+}(z,t)$. Here, functions $\beta_{1}(\tilde{\xi})$ and $\beta_{2}(\tilde{\xi})$ have the same form as in Eq. (\ref{betas}). However, $\tilde{\xi}$ is defined differently; $\tilde{\xi}(\delta k)\equiv2[\Delta \omega_{s}(\delta k)-\omega_{ac}]/\gamma\Delta B_{0}$. Note that, besides the shift of $\Delta \omega_{s}$ on the frequency $\omega_{ac}$, there is a factor $2$ in $\tilde{\xi}$. The reason why we have $\gamma\Delta B_{0}/2$ in $\omega_{R}$ (instead of $\gamma\Delta B_{0}$ as in $\omega_{\downarrow\uparrow}$ for the dc case) is that the ac modulation splits into the rotating and counter-rotating parts, and only the contribution from the rotating component has to be taken into account within the RWA (see SM for more details).

\begin{figure}[htp] \centerline{\includegraphics[clip, width=1 \columnwidth]{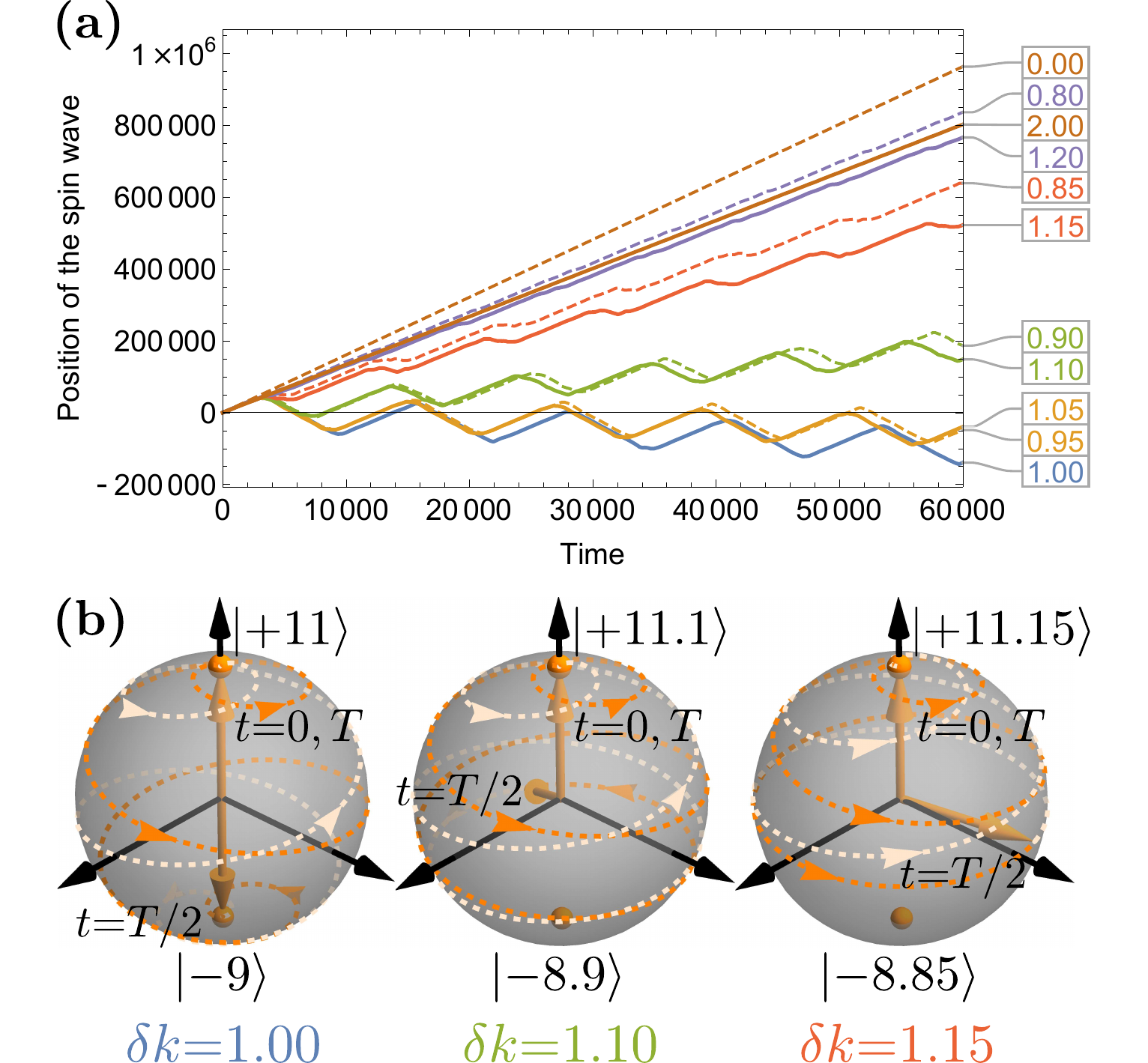}}
	
	\protect\caption{(a) The time-dependent position of the spin waves with different wave vectors $q/2+\delta k$ after the ac modulated magnonic crystal with $\omega_{ac}=0.00316$ is activated at $t=0$. Other parameters here are the same as in Fig. \ref{fig:to_and_fro_vs_BS_DC}. (b) Trajectories of the spin waves with different $\delta k$ near the \emph{shifted resonance} on the Bloch sphere. Here, the period $T=2\pi/\omega_{R}(\tilde{\xi})$ is different from the dc case.}
	
	\label{fig:to_and_fro_vs_BS_AC}
\end{figure}

By properly tuning $\omega_{ac}$, the resonant spin-wave wave vector can be \emph{noticeably} shifted from $q/2$. As one may observe from Fig. \ref{fig:to_and_fro_vs_BS_AC}(a), if $\omega_{ac}$ is set to be $0.00316$, only the spin waves around $k=11$ perform the to-and-fro motion accurately enough. For the chosen parameters, namely, $A=2$ and $q=20$, one has $\Delta\omega_{s}(\delta k=1)\approx0.00316$ and, therefore, the \emph{shifted resonance} indeed happens at $\tilde{\xi}\approx0$.

Furthermore, in Fig. \ref{fig:to_and_fro_vs_BS_AC}(a), the period of the zigzag curves around the horizontal axis is doubled as compared with the one in Fig. \ref{fig:to_and_fro_vs_BS_DC}(a). In addition, the width of the $\delta k$ intervals at which the to-and-fro motion develops becomes approximately half of the one in the dc case. For example, in Fig. \ref{fig:to_and_fro_vs_BS_AC}(a), the to-and-fro motion can be observed within the interval $10.85\lesssim k\lesssim11.15$, while for the dc magnonic crystal presented in Fig. \ref{fig:to_and_fro_vs_BS_DC}(a) it develops when the spin-wave wave vectors are within $9.67\lesssim k\lesssim10.33$. All these observations are consistent with our understanding based on the solutions presented by Eqs. (\ref{sol_Mx_static}) and (\ref{sol_Mx_time_depen}).

\noindent\emph{Bloch-sphere representation.} The spin-wave solutions, e.g., the RWA solution in the ac case [cf. Eq. (\ref{sol_Mx_time_depen}) and Eq. (S20) in SM], can be written as
\begin{align}\label{Bloch_sphere}
	\ket{\text{SW}}\equiv&S^{+}(z,t)
	\nonumber
	\\
	\equiv&e^{-i\varphi_{g}(t)}\bigg\{\bigg[\cos(\frac{\omega_{R}t}{2})-i\beta_{1}(\tilde{\xi})\sin(\frac{\omega_{R}t}{2})\bigg]\ket{k}
	\nonumber
	\\
	&+\beta_{2}(\tilde{\xi})\sin(\frac{\omega_{R}t}{2})e^{i(\omega_{ac}t+\varphi_{ac}-\pi/2)}\ket{k-q}\bigg\}.
\end{align}
Here, $\ket{k}\equiv(\Delta S)e^{i(k_{+}z)}$ and $\ket{k-q}\equiv(\Delta S)e^{i(-k_{-}z)}$ are two states associated with the wave vectors $k=k_{+}$ and $k-q=-k_{-}$. In the second line, the global phase is given with $\varphi_{g}(t)\equiv\omega_{ac}t/2+\Omega t-\varphi_{s}$. [For the dc case, one can obtain the same result as in Eq. (\ref{Bloch_sphere}) by replacing $\omega_{R}$ with $\omega_{\downarrow\uparrow}$ and taking both $\omega_{ac}$ and $\varphi_{ac}$ to be $0$.]

The state $\ket{\text{SW}}$ described by Eq. (\ref{Bloch_sphere}) can be represented by a vector, whose ending point is moving on the surface of a Bloch sphere with $\ket{k}$ and $\ket{k-q}$ to be its north and south poles. The motion of the states with different initial wave vectors $k$ on the Bloch spheres, after turning on the dc and ac magnonic crystal, are shown in Figs. \ref{fig:to_and_fro_vs_BS_DC}(b) and \ref{fig:to_and_fro_vs_BS_AC}(b), respectively. As demonstrated by Fig. \ref{fig:to_and_fro_vs_BS_DC}(b), at the resonance (i.e., when $\delta k=0$), during the first period of the to-and-fro motion, the spin-wave state, which is initially located at the north pole, starts to move to the south pole, and then, after passing the south pole, returns to the north pole. Note that, the spin wave is propagating forward when the state is on the north hemisphere, and vice versa. However, at $\delta k=0.2$, the trajectory does not pass through the south pole, and its portion below the equator becomes less than the one above the equator. Finally, when $\delta k=\delta k_c\approx0.33$, the full trajectory is on the north hemisphere only, which indicates the disappearance of the to-and-fro motion.

Figure \ref{fig:to_and_fro_vs_BS_AC}(b), plotted for an ac magnonic crystal with $\omega_{ac}=0.00316$, shows a more complicated behavior, which includes a non-trivial precession of the states on the Bloch spheres. In the presented case, the resonant spin-wave wave vector is shifted to $k\approx11$ (and $9$).

\noindent\emph{Discussion and outlook.} (i) The dc or ac magnonic crystal created by a metallic meander structure (which is the ``hardware" part of the magnonic crystal) with a fixed period $d=2\pi/q$ can be utilized for the control of the spin waves around $k=q/2$ or where $k$ satisfies the shifted resonance condition $|\omega_s(k)-\omega_s(k-q)|=\omega_{ac}$.  As it was presented in Figs. \ref{fig:to_and_fro_vs_BS_DC} and \ref{fig:to_and_fro_vs_BS_AC}, the spin waves, which are at the resonances (the regular or shifted ones), perform the to-and-fro motion, while the out-of-resonance spin waves are almost unaffected. The to-and-fro frequency is determined by the band gap, which is controlled by the amplitude of the supplied dc or ac current. The band gap is $\gamma\Delta B_{0}$ or $\gamma\Delta B_{0}/2$ (notice the factor $1/2$ here) in the dc or ac case. The dc spin-wave dynamics has been demonstrated in Ref. [\onlinecite{karenowska2012oscillatory}] as the oscillatory energy exchange between the wave and its counter-propagating reflection. However, the ac dynamics still remains to be investigated experimentally.

\begin{figure}[htp] \centerline{\includegraphics[clip, width=1 \columnwidth]{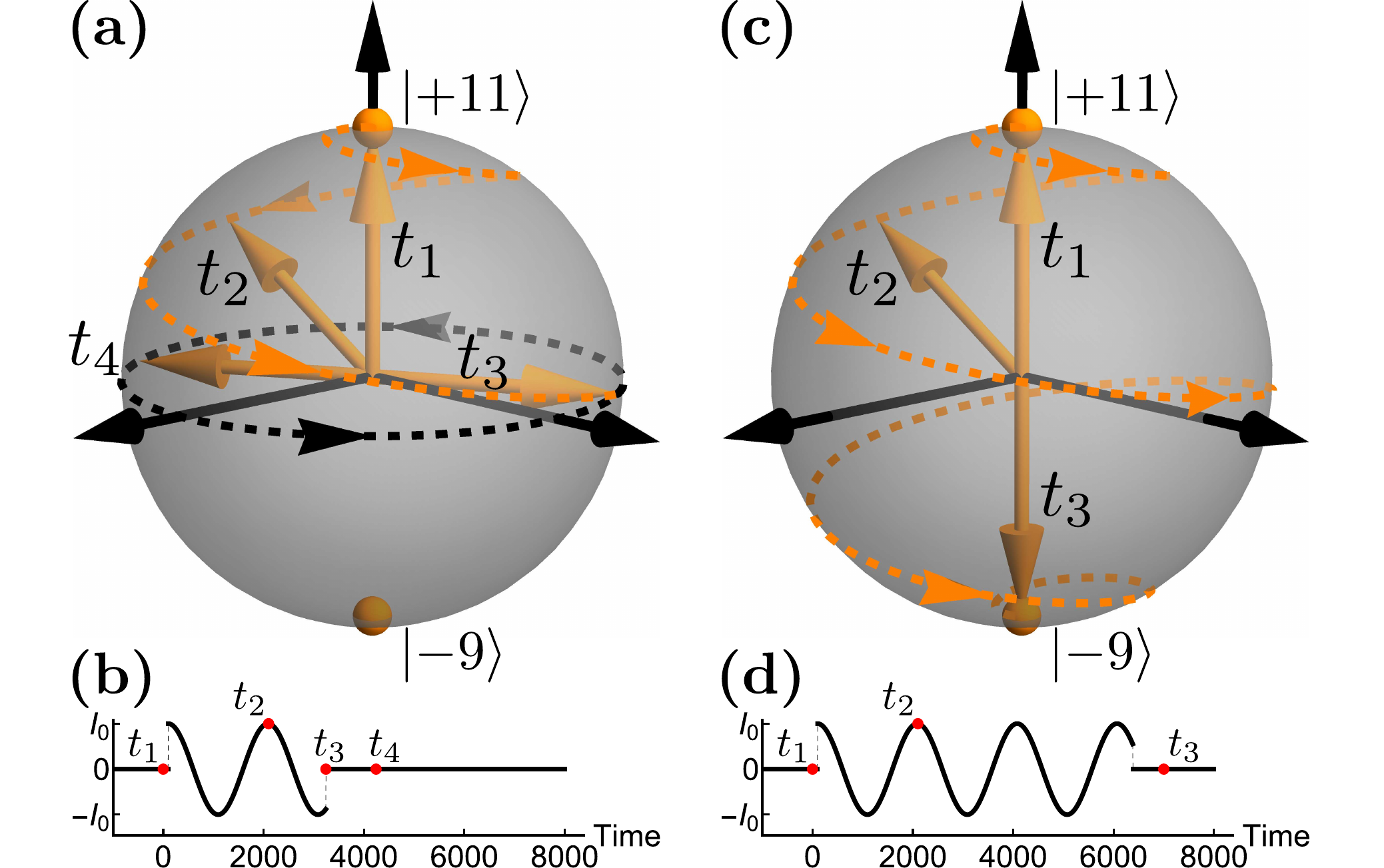}}
	
	\protect\caption{(a) Motion of the spin wave with the initial wave vector $k=11$ on the Bloch sphere when sending an ac $\pi/2$ pulse as shown in (b). The dashed orange curve with arrowheads is the trajectory during the activation of the $\pi/2$ pulse, while the dashed black curve on the equator indicates motion after $t_{3}$. (c) Motion when sending $\pi$ pulse (d). Both pulses are generated at $\omega_{ac}=0.00316$. The widths of the $\pi/2$ and $\pi$ pulse are $\pi/2\omega_{R}$ and $\pi/\omega_{R}$, where $\omega_{R}$ is determined by the intensity of the pulse.}
	
	\label{fig:pulses}
\end{figure}

(ii) When the spin wave is at the exact shifted resonance (i.e., $\tilde{\xi}=0$), Eq. (\ref{Bloch_sphere}) is reduced to $\ket{\text{SW}}=e^{-i\varphi_{g}(t)}\allowbreak\left\{\cos\left[\theta(t)/2\right]\ket{k}+\sin\left[\theta(t)/2\right]e^{i\phi(t)}\ket{k-q}\right\}$. Here, $\theta(t)\equiv\omega_{R}t$ and $\phi(t)\equiv\omega_{ac}t+\varphi_{ac}-\pi/2$. This state can be considered as a spin-wave ``qubit'', which is characterized by the polar angle $\theta(t)$ and azimuthal angle $\phi(t)$. Recall that $\omega_{R}(\tilde{\xi}=0)=\gamma\Delta B_{0}/2\propto I_{0}$; therefore, the time evolution of this ``qubit'' [i.e., $\theta(t)$ and $\phi(t)$] is fully controlled by the intensity and the ac modulation of the magnonic crystal. Consequently, one can manipulate this spin-wave based macroscopical ``qubit'' using a properly designed ac pulse. For example, Fig. \ref{fig:pulses}(a) demonstrates that one can bring the state from the north pole to the equator on the Bloch sphere utilizing a $\pi/2$ pulse shown in Fig. \ref{fig:pulses}(b). It is also possible to flip the spin-wave ``qubit'' by the $\pi$ pulse, see, e.g., Figs. \ref{fig:pulses}(c) and \ref{fig:pulses}(d), and more discussions in SM. Note that the spin-wave ``qubit'' controlled by the ac magnonic crystal has a nonzero energy mismatch $\Delta\omega_{s}=\omega_{ac}$. This leads to a time-dependent $\phi(t)$ and, thereby, opens a way to manipulate the azimuthal angle of this ``qubit'' and makes all single-qubit operations possible \cite{nielsen2002quantum}.

(iii) Finally, we discuss the feasibility of the proposed scheme. To make the RWA valid, $\omega_{ac}$ has to be much greater than $\gamma\Delta B_{0}$. In addition, to minimize the higher-order effects from the spin-wave states with wave vectors $k+q$, $k\pm2q$, $\cdots$, the inequalities of the kind $||\omega_{s}\left(k+q\right)-\omega_{s}\left(k\right)|-\omega_{ac}|\gg\gamma\Delta B_{0}/2$ need to be fulfilled. In particular, one may require $|\omega_{s}(q)-\omega_{s}(0)|-\omega_{ac}\gg\gamma\Delta B_{0}/2$. Let us take the experimental parameters in Ref. [\onlinecite{karenowska2012oscillatory}] as an example. In this measurement, the value of $\gamma\Delta B_{0}$ was $2\pi\times11$ MHz, while $|\omega_{s}(q)-\omega_{s}(0)|$ was roughly $2\pi\times100$ MHz. In this case, an ac modulation with $\omega_{ac}$ around $2\pi\times50$ MHz would be suitable for the experimental studies of the shifted resonance. Spin-wave attenuation and decoherence are the two other factors that need to be considered. Let us estimate the operation time required to bring the spin-wave state from the north to the south pole. Using the same value of $\gamma\Delta B_{0}$, we find this time to be $2\pi/\gamma\Delta B_{0}\approx91$ ns, which is shorter than both lifetime and coherence time of spin waves in YIG (cf. Refs. \cite{serga2010yig,collet2017spinwave,mahmoud2020introduction}). In the end, we estimate the length of the device. In practice, the slopes of the curves plotted in Figs. \ref{fig:to_and_fro_vs_BS_DC}(a) and \ref{fig:to_and_fro_vs_BS_AC}(a) are determined by the group velocities of the incident and reflected spin-wave packets, which may vary dramatically for different magnetic material and geometries. In the case of the magnetostatic spin waves studied in Ref. [\onlinecite{karenowska2012oscillatory}], the spin-wave group velocity $|v_g|$ was found to be about $27404$ m/s. Such a spin-wave packet would perform to-and-fro motion with maximum displacement $|v_{g}|\times2\pi/\gamma\Delta B_{0}\approx2.5$ mm, and therefore, the meander structure must have a size of at least $2.5$ mm in order to confine this bouncing packet. In principle, the operating range of $\omega_{ac}$ can be enlarged to a GHz range by fabricating a meander structure with a shorter period $d$, and the operation time can be further reduced by applying a bigger $I_{0}$.

With the use of the techniques accessible nowadays, the \emph{tunable} ac magnonic crystal can be exploited for controlling spin waves with different wave vectors. One can excite and detect many spin waves with different frequencies in a magnetic sample at the same time through one antenna setup \cite{scalability}. The tunability of the ac magnonic crystal allows one to use a single hardware (the meander pattern) to simultaneously manipulate multiple spin-wave pairs in one waveguide. Each pair formed by waves with different frequencies is independently operated by a suitable ac pulse that satisfies the \emph{shifted resonance}. As a result, the \emph{tunable} ac magnonic crystal can serve as a new building block of the computing device studied in Ref. [\onlinecite{balynsky2021quantum}] and improve its spatial scalability with no compromise on computational time.

\noindent\emph{Acknowledgements.} We appreciate Artem Abanov for reading the manuscript. We thank Karen Michaeli and Alexey Belyanin for discussions. We gratefully acknowledge Dr. Alexander Khitun for the interest in this paper as well as providing us with useful comments on his work.

\bibliography{MyBIB}

\end{document}